\documentclass[%
 reprint,
 amsmath,amssymb,
 aps,
]{revtex4-2}

\usepackage{placeins}
\usepackage{comment}
\usepackage{caption}
\usepackage{latexsym}
\usepackage{amssymb}
\usepackage{amsmath}
\usepackage{subcaption} 
\usepackage{graphicx}
\usepackage{dcolumn}
\usepackage{bm}
\begin{document}


\title{Growth of a viscoplastic blister underneath an elastic sheet}

\author{Torstein Sæter$^1$}
\author{Olivier Galland$^2$}
\author{Blandine Feneuil$^{1,3}$}
\author{Andreas Carlson$^1$}%
 \email{acarlson@math.uio.no}
\affiliation{$^1$
Department of Mathematics, Mechanics Division, University of Oslo, Oslo 0316, Norway}%
\affiliation{$^2$
Physics of Geological Processes, The Njord Centre, Department of Geosciences, University of Oslo, Blindern, 0316, Oslo, Norway}
\affiliation{$^3$
Current address: SINTEF Industry, Petroleum department, Trondheim, Norway.}
\date{\today}

\begin{abstract}
Inspired by the formation of geological structures as earth's crust deforms by magmatic intrusions, we investigate the elastohydrodynamic growth of a viscoplastic blister under an elastic sheet. By combining experiments, scaling analysis and numerical simulations we reveal a new regime for the growth of the blister's height $\sim t^{5/9}$ and radius $\sim t^{2/9}$. A plug like flow inside the blister dictates its dynamics, whereas the blister takes a quasi-static self-similar shape given by a balance in the pressure gradient induced by bending of the elastic sheet and the fluid's yield stress.
\end{abstract}

\maketitle

The intrusion of a liquid underneath an elastic sheet resting on a non-deformable solid has been deployed as a generic model to understand the formation of geological structures such as magmatic sills and laccoliths \cite{POLLARD1973311,  doi:10.1029/2010JB008108, doi:10.1029/91JB00600}, and ice sheet relaxation \cite{lai2021hydraulic}. In models of \textit{e.g.,} sill and laccolith emplacement, the intruding magma is commonly assumed to be Newtonian, and the solid is assumed to deform by elastic bending and tensile fracturing. However, there is growing evidence that some magmas exhibit highly non-Newtonian behaviour \citep{BalmforthViscoplastic,CARICCHI2007402,cordonnier2012}, such that the magma can transport as plug flow \citep{morgan2008}. Conversely, rock layers along which magma is emplaced can exhibit complex viscoplastic rheological response to magma flow \cite{doi:10.1002/2016JB013754,GALLAND2019120}. To which extent these non-linear phenomena affect the dynamics of these geophysical systems is unclear.

In this context, the dual nature of a viscoplastic fluid, i.e. when the applied stress is above the yield stress $\tau_0$ it flows like a liquid and below $\tau_0$ it is a solid, is particularly interesting to capture the consequences of the fluid-and-solid nature of the flowing matter. We describe in here the spatiotemporal growth of a viscoplastic blister underneath an elastic sheet.

When an elastic sheet is resting on a solid pre-wetted by a fluid film, it has been shown that the height and radius of the formed blister is determined by local effects at the intrusion tip, which couple to the interior quasi-static solution of the shape of the sheet \cite{PhysRevLett.111.154501}. Other effects have also been demonstrated to affect the blister's dynamics, e.g. vapour at the intrusion tip \cite{wang2018tip}, adhesion to the supporting solid \cite{Ball2018,lister_skinner_large_2019, hosoi2004peeling}, gravitational flow \cite{hewitt_balmforth_debruyn_2015}, the height of the pre-coated film layer \cite{PhysRevFluids.4.124003} as well as stochastic effects \cite{carlson2018fluctuation}. The influence of a deformable elastic sheet on the resulting flow also has been shown in other settings such as the suppression of viscous fingering in a Hele-Shaw geometry \cite{juel_2018} and the effect of the porous-base \cite{lai2021hydraulic, Chase} . 

Fluid models that account for viscoplasticity have been used to provide a more realistic representation of mud, glacier and magma flows on the Earth's surface \cite{liu_mei_1989,BalmforthViscoplastic,doi:10.1146/annurev.earth.31.100901.141352}. Recently, viscoplastic fluids have also attracted much interest in interfacial flows, demonstrated to affect coating, drop spreading and coalescence  \cite{PhysRevLett.123.148002,jalaal_stoeber_balmforth_2021, sanjay_lohse_jalaal_2021}.

\begin{figure}[!htbp]
\includegraphics[scale = 0.2]{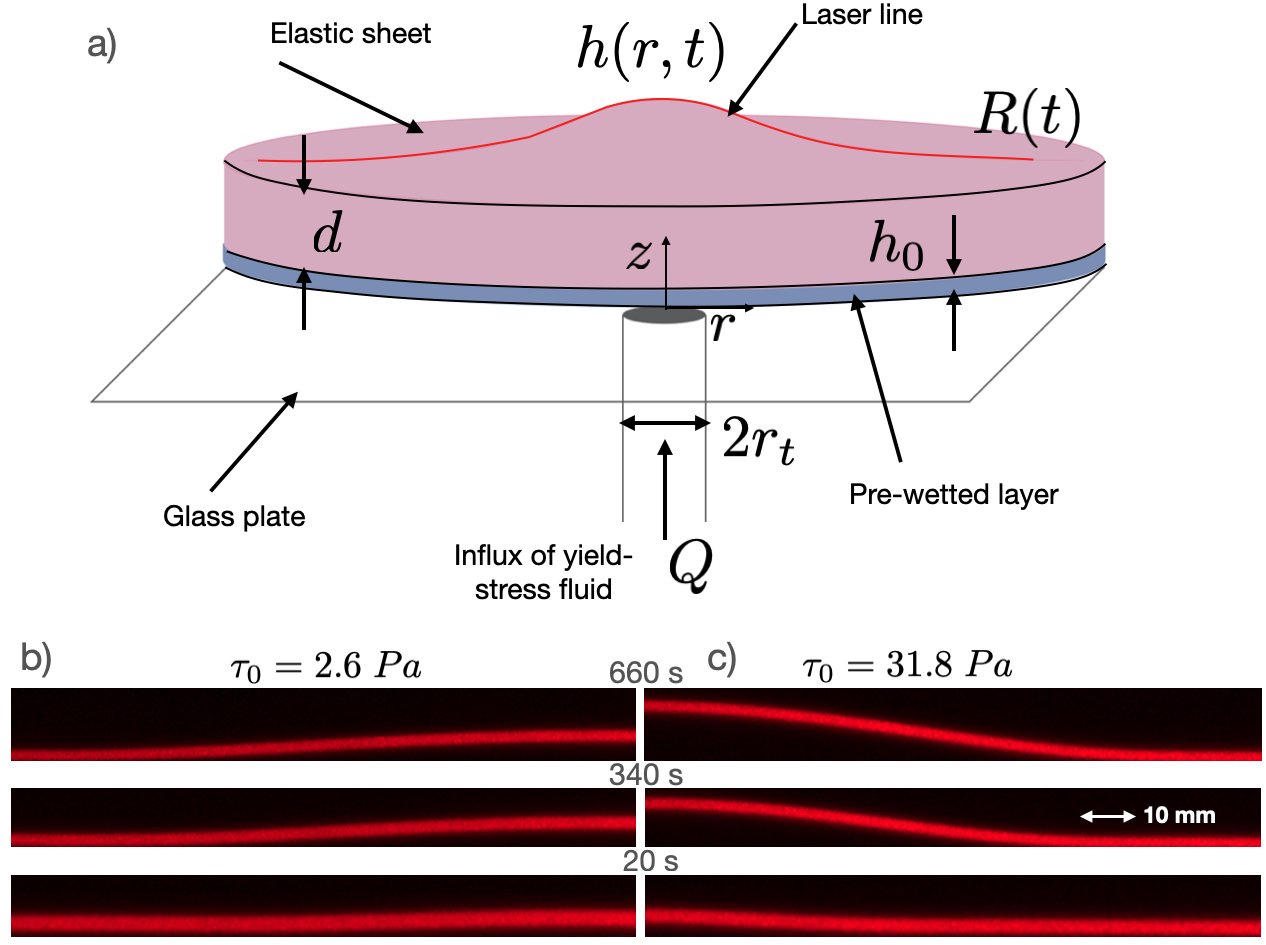}
\caption{\label{fig:intro13} a) Schematic drawing of the studied system. An elastic sheet, of Young's modulus $E = 0.126\;$ MPa, thickness $d = 10 \;$mm, and Poisson's ratio $\nu = 0.5$, rests on top of pre-wetted layer of yield stress fluid Carbopol, of thickness $h_0$. Solutions of the same Carbopol as that of pre-wetted layer is injected at constant $Q$ beneath the elastic sheet through a small tube of radius $r_t = 2\;$mm. The resulting blister is measured by tracking the laser-line and extracting the height profile, $h(r,t)$. Experimental profiles $h(r,t)$ of the sheet at $t=[ 20, 340, 660]$ s when injecting a fluid with b) low ($\tau_0=2.6\;Pa$) and c) high ($\tau_0=31.8\;Pa$) yield stress.}
\end{figure}
\normalsize

We combine here the effects of having an elastic interface that can bend and look at the formation and growth of a blister as we inject a viscoplastic fluid. Intuitively, we might suspect that the large stress at the peeling front, as predicted from the Newtonian case \cite{PhysRevLett.111.154501}, will locally fluidize the material, but how is the peeling front dynamics affected by the yield stress threshold? To answer this, we make experiments adapting the design from previous Newtonian elastohydrodynamics studies \cite{PhysRevLett.111.154501,michaut2019}. Fig. \ref{fig:intro13}a shows a schematic representation of the experimental system, where we use Carbopol as a model yield-stress fluid as it has well characterised rheological properties and small thixotropic effects \cite{COUSSOT201431}. We prepare four different Carbopol solutions, giving a range in yield stresses $ \tau_0 \cap [2.6-47.2]$ Pa when fitting the {Herschel-Bulkley} model to the rheological measurements, see Supplemental Material (SM) for details.

A solid plate is coated by a thin layer of Carbopol with height $h_0$  $\approx [0.15-0.35]$ mm. The elastic sheet initially rests upon the Carbopol layer, before injecting Carbopol with a controlled flux $Q\cap[1.7,3.3,6.7]\cdot10^{-7}m^3/s$ through the tube at the center of the plate. The sheet is made of a silicone based elastomer (Zhermack, Elite Double) with a Young's modulus $E=0.126$ MPa. A laser line is used to visualise the sheet's deflections $h(r,t)$ as the blister grows in time and space by using side view imaging.

The spatiotemporal growth of the blister for two Carbopol solutions, $\tau_0=2.6$ Pa and $\tau_0=31.8$ Pa, are shown in Fig. \ref{fig:intro13}b-c, respectively. There are some noteworthy features to highlight in Fig. \ref{fig:intro13}b. A distinct effect of $\tau_0$ on the growth of the blister's radius $R(t)$ and its height at the center $h(0,t)$ is observed. The ratio between the deflection and the radius remains small throughout the experiments ($h(0,t)/R(t) \leq 0.1$), and the flow is highly viscous with a small Reynolds number. Our measurements of the blister profiles show a high degree of axis-symmetry around the tube where Carbopol is injected in the experiments.

To give a theoretical description of the elastohydrodynamics of the viscoplastic intrusion observed in Fig. \ref{fig:intro13}b, we use the small aspect ratio and the viscous flow allowing us adopt the axisymmetric lubrication theory. There is a minor effect from gravity (illustrated in the SM), which we have neglected here for clarity, an assumption strictly valid for lengths below the elastogravity length $L_{eg} = \big(\frac{B}{ \rho g}\big)^{1/4}\approx 0.035$ m, with $\rho$ the liquid density and $g$ gravitational acceleration. The bending modulus of the sheet is fixed $B = \frac{Ed^3}{12(1-\nu^2)}= 0.014$ N$\cdot$m, with $d$ the sheet thickness and $\nu=0.5$ the Poisson's ratio. We use the fact that there are only small deformations of the sheet and we can then neglect in-plane tension \cite{Elasticity}. Thus the bending pressure is described as $p(r,t) = B \nabla^4 h(r,t) $ for small slopes. To capture the principal viscoplastic effect we use the Bingham model, which can be combined with the integrated lubrication equations to derive the expression for the vertical yielding line $Y(r,t)$ of the fluid, where the shear stress $\tau_{rz}$ equals the yield stress $\tau_0$ \cite{liu_mei_1989,BalmforthViscoplastic, BALMFORTH2006103},  
\begin{equation}
Y(r,t) = max\bigg(0,\frac{h(r,t)}{2} - \frac{\tau_0}{\mid \frac{\partial p(r,t)}{\partial r} \mid}\bigg). 
\label{eq:yield_limit}
\end{equation}
We expect the shear profile $\tau_{rz}$ to be symmetric, and will therefore have another corresponding yielding position along $z(r,t)= h(r,t)-Y(r,t)$. Between the two yielding lines we have a plug flow, which flows with a constant speed and zero shear rate \cite{Balmforth1999ACT}.

We follow the standard procedure to derive the thin film equation, where an additional term including $Y(r,t)$ appear from the viscoplastic Bingham model \cite{oron1997long,liu_mei_1989, BalmforthViscoplastic,BALMFORTH2006103}:
\begin{equation}
\label{eq:evolution}
\begin{split}
\frac{\partial h(r,t)}{\partial t} = &\frac{\partial}{6\mu r\partial r}\bigg(rY(r,t)^2(3h(r,t)-2Y(r,t))\frac{\partial p(r,t)}{\partial r} \bigg)\\ + &w(r).
\end{split}
\end{equation}
We have here introduced $w(r) = \frac{2Q}{\pi r_t^2}\bigg(1- \big(\frac{r}{r_t}\big)^2\bigg)$ for $r\leq r_t$, which is the Poisseuille flow through the tube with radius $r_t$ allowing us to prescribe the influx $Q$. Note that a no-slip condition has been applied at both surfaces and that $\mu$ is the effective fluid viscosity. Equation \eqref{eq:evolution} has two particularly interesting limits. If $Y(r,t)=h(r,t)$ Newtonian flow is recovered and predicts the blister to grow with a height $\sim t^{8/22}$ \cite{PhysRevLett.111.154501}. As $Y(r,t)\rightarrow0$ we expect a plug flow to form and the flow to be dominated by the effects from $\tau_0$. 

\begin{figure}[h!]
\centering
\includegraphics[scale = 0.25]{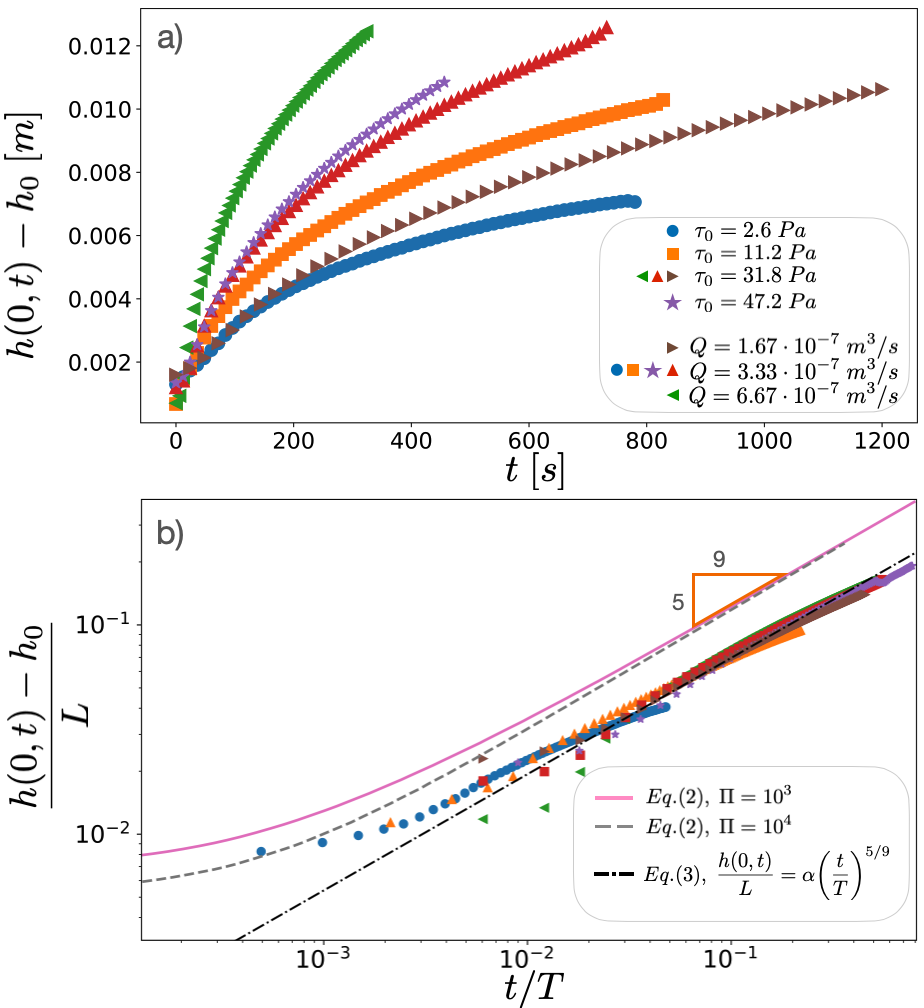}
\caption{\label{fig:heightinTime} a) Plots of experimental results for the blister height $h(0,t)-h_0$ for different yield stress $\tau_0$ and flux $Q$. b) Results from a) scaled by the length $L = \big({B}/{\tau_0}\big)^{1/3}$ and the time $T = {B}/{Q\tau_0}$ while comparing with the numerical simulations of the non-dimensional version of equation \eqref{eq:evolution}. The dashed-dotted line shows the derived scaling law of equation \eqref{eq:height_powerlaw}, where we have determined $\alpha$ = 0.25.}
\end{figure}
\begin{figure*}[!htbp]
\includegraphics[scale = 0.35]{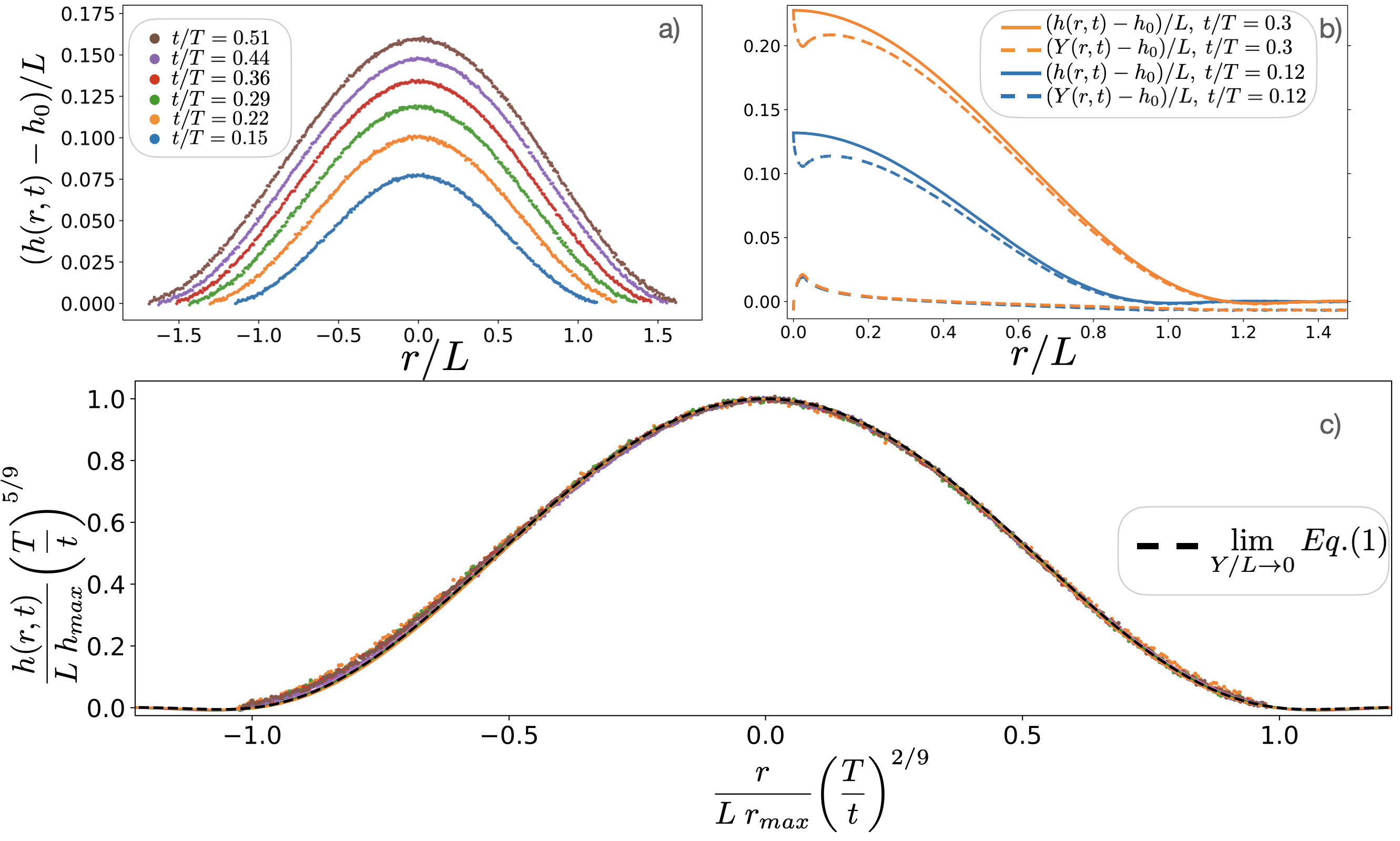}
\caption{\label{fig:profiles} a) Plot of laboratory blister profiles at different times with $\tau_0 = 47.2\; Pa$ and $Q=3.3\cdot10^{-7}\;m^3/s$. b) Plots of numerical blister profiles for two times, computed from non-dimensional version of equation \eqref{eq:evolution} with $\Pi = 1000$. Dashed lines for each profile represent the non-dimensional yield limits $Y(r,t)/L$ and $(h(r,t)-Y(r,t))/L$; plug flow occurs in-between the yield lines. In c) we scale our experimental and numerical height profiles from a) and b) (11 profiles between the times $t/T=0.12-0.3$) by using the scaling laws in equation \eqref{eq:height_powerlaw} and equation \eqref{eq:radius_powerlaw}. The profiles fall onto a self-similar shape, described by the solution of the non-dimensional form of equation \eqref{eq:yield_limit} when $Y/L \to 0$, where we have $L = \big({B}/{\tau_0}\big)^{1/3}$, $T = {B}/{Q\tau_0}$, $h_{max} = \frac{h(0,t)T^{5/9}}{Lt^{5/9}}$ and $r_{max} =\frac{R(t)T^{2/9}}{Lt^{2/9}} $.}
\end{figure*}

Before moving on to the experiments and the numerical simulations of equation \eqref{eq:evolution}, we derive a scaling law for the spatiotemporal blister growth. If we scale the height $h(r,t)$ with $h(0,t)$ and the radial direction $r$ with $R(t)$, while assuming unidirectional flow from $r=0$ and outwards, in the limit $Y(r,t)\rightarrow 0$, equation \eqref{eq:yield_limit} gives $\frac{h(r,t)\partial p(r,t)}{2\partial r} \sim \frac{Bh^2(0,t)}{R(t)^5}\sim(h(0,t)/2\tau_0)$. We impose mass conservation, where the volume $V= \int_0^{R(t)}2\pi h(r,t)r dr\sim h(0,t)R(t)^2 \sim Qt$, implicit in equation \eqref{eq:evolution}, which gives the scaling laws for the blister height $h(0,t)$ and radius $R(t)$: 

\begin{equation}
h(0,t) \sim \bigg(\frac{4\tau_0^2Q^5}{B^2}\bigg)^{1/9}t^{5/9} \sim LT^{-5/9}t^{5/9}
\label{eq:height_powerlaw}
\end{equation}
\begin{equation}
R(t) \sim \bigg(\frac{Q^2B}{2\tau_0}\bigg)^{1/9}t^{2/9} \sim L T^{-2/9}t^{2/9}
\label{eq:radius_powerlaw}
\end{equation}

A length $L = (\frac{B}{\tau_0})^{1/3}$ and time $T = \frac{B}{Q\tau_0}$ appear in equation \eqref{eq:height_powerlaw} and equation \eqref{eq:radius_powerlaw}, which characterise our system. 

We make equation \eqref{eq:yield_limit} and equation \eqref{eq:evolution} non-dimensional by scaling $h(r,t)$ and $r$ with $L$ and time $t$ with $T$, giving only one non-dimensional number appearing in front of the first term on the right-hand-side of equation \eqref{eq:evolution} i.e. $\Pi = \frac{B}{6\mu Q}$, which gives the ratio between elastic and viscous forces. We assume the effective viscosity in our experiments to be set by the shear rate in the regions where the fluid is yielded, because the flows here are determining our viscous timescale. We estimate the shear rate by assuming a parabolic velocity profile for $z > h(r,t) - Y(r,t)$ and $z<Y(r,t)$, and obtain $\dot{\gamma}_{max} \approx \frac{\partial u}{\partial z}|_{z=0, z = h}  \sim \frac{Q}{R(t)Y(r,t)^2}$. Setting $Q\approx10^{-7}\;m^3$/s, $ R(t)\approx 10^{-1}\;$ m, and $Y(r,t)\approx 10^{-4}$ m gives $\dot{\gamma}_{max} \approx 100 \;s^{-1}$. This is a conservative estimate for the shear rate and we therefore assume $\dot{\gamma}\approx[10-100]s^{-1}$ in the experiments. From the rheological viscosity curves (see SM), we estimate the effective viscosity $\mu\approx[0.1-10]$ Pa$\cdot$s, which translates to $\Pi\approx[10^2-10^4]$. 

We use the blister's maximal height $h(0,t)$ as a metric to characterise the dynamics to test the predictions from equation \eqref{eq:height_powerlaw} and equation \eqref{eq:radius_powerlaw}. The bending stiffness $B$ is fixed, while we vary the yield stress $\tau_0$ and the flux $Q$ in the experiments shown in Fig. \ref{fig:heightinTime}a. It is clear that $h(0,t)$ is affected by both parameters by steepening of the blister profile. With these experiments at hand, the experimental data can be rescaled by $h(0,t)/L$ and $t/T$, as shown in Fig. \ref{fig:heightinTime}b, and collapse onto a single curve following the proposed scaling law of equation \eqref{eq:height_powerlaw}. At late times in the experiments, gravity is starting to affect the dynamics and we enter a transitional regime where bending and gravitational forces become comparable. 

The experiments are complemented by numerical solutions of equation \eqref{eq:evolution}, which is discretised by linear finite elements and the solution method is described in \cite{PhysRevFluids.4.124003}. We use an adaptive time-stepping routine, with a time step limit of $\Delta t/T = 9\cdot 10^{-6}$, and a discretisation in space $\Delta r/L \in [0.0005-0.005]$. We have defined $Y(r,t)=max\bigg(\epsilon,h(r,t)/2-\frac{1}{\mid\frac{\partial}{\partial r} \nabla^4h(r,t)\mid}\bigg)$, with a regularisation parameter $\epsilon = 10^{-6}$ similar to \citep{jalaal_stoeber_balmforth_2021}, and we tested that the results are unaffected by this choice of $\epsilon$. The simulations are started with a pre-wetted layer $h_0/L =0.0069$, and with $Y(r,t)=0$, which is gradually introduced until $t/T = 10^{-4}$. The results from the numerical simulations are shown with the experiments in Fig. \ref{fig:heightinTime}b, they are fairly insensitive to the non-dimensional number $\Pi$, and recover the scaling law in equation \eqref{eq:height_powerlaw}. The small time shift separating the experiments and the simulations (a factor of $2.5$ in time) may stem from the minimal rheological model we use to describe Carbopol \cite{COUSSOT201431}, wall slip or the simplified influx condition. Regardless, the simulations overall represent well the elastohydrodynamic growth of the viscoplastic blister.



Next, we want to understand the interface dynamics of the blister and if it adopts a self-similar shape. Fig. \ref{fig:profiles}a shows the blister shape in time from the experiments. To better understand the viscoplastic flow in the blister, we plot in Fig. \ref{fig:profiles}b the non-dimensional height profile and the yielded lines for two time instances from a numerical simulation. It is clear that a major part of the area is encompassed between the yielding lines (dashed lines) in Fig. \ref{fig:profiles}b with a significant plug flow, whereas a flow with a viscous gradient is limited to thin areas near the two bounding solid surfaces. This further supports our argument allowing us to derive equation \eqref{eq:height_powerlaw} and equation \eqref{eq:radius_powerlaw}. To go one step further, we now rescale the height profiles from the experiments and the numerical simulations data (eleven profiles are shown in-between the two time-points in Fig. \ref{fig:profiles}b, using the predictions from equation \eqref{eq:height_powerlaw} and equation \eqref{eq:radius_powerlaw}, where we normalise each profile with: $h_{max} = \frac{h(0,t)T^{5/9}}{Lt^{5/9}}$ and radial value, $r_{max} =\frac{R(t)T^{2/9}}{Lt^{2/9}} $, see SM. The experimental and numerical data fall onto a profile for a universal blister shape, see Fig. \ref{fig:profiles}c. 

The self-similar blister shape can be understood if we look back to equation \eqref{eq:yield_limit} in the limit $Y(r,t)\rightarrow 0$ that reduces to a non-linear ordinary differential equation (ODE): $h(r)/ d\nabla^4 h(r)/dr= 2\tau_0 sgn(h_0-h)$ \citep{jalaal_stoeber_balmforth_2021}. By scaling $h(r)$ and $r$ with $L$, equation \eqref{eq:yield_limit} becomes parameter free. To determine the blister shape we use a numerical profile from our simulation, $\Pi = 1000$ and $t/T = 0.75$, as an initial guess. We need five boundary conditions to define our problem. We choose three conditions at the centerline, where we pin the height to the maximum height of the simulation profile, $h_{max}$, ${h(0)} = h_{max}$, and set two symmetry conditions, $\frac{\partial {h}(0)}{\partial r} = 0$ and $\frac{\partial^3 {h}(0)}{\partial r^3} = 0$. We supplement with two boundary conditions at $\tilde{R}=1.3 \times R$, where we set ${h}(\tilde{R}) = h_0$ and $\frac{\partial^2 {h}(\tilde{R})}{\partial r^2} = 0$. The solver \textit{Solve\_bvp} retrieved from the scipy package in Python is used to solve the ODE. In Fig. \ref{fig:profiles}c, the solution for the height profile is shown together with the experiments and the numerical simulations of equation \eqref{eq:evolution}, which gives a perfect agreement for the predicted shapes. 

Intrusion of flowing matter with complex rheological properties underneath an elastic layer or inside an elastic matrix appear in a myriad of geological processes. Both non-linear magma plug flow \citep{morgan2008} and viscoplastic deformation of magma intrusion's host rocks \citep{GALLAND2019120} have been documented to play a key role in magma emplacement in the Earth's crust. Our study quantifies the dynamics of a viscoplastic blister growth underneath an elastic sheet, and evidences a viscoplastic regime for the growth of its height and radial expansion. The blister is shown to adopt a self-similar quasi-static shape in time that is set by the balance between the pressure gradient induced by the bending of the elastic sheet and the yield stress of the intruding fluid. These results suggest that the growth of geological structures such as sills and laccoliths exhibit distinct dynamics than those predicted from a Newtonian model if the intruding fluid can support a critical stress before starting to flow. Our work gives a first look at the elastohydrodynamics of a growing viscoplastic blister, but  highlight that there are many more questions in these configurations that remain to be understood such as additional non-Newtonian effects in the bulk and the elastic sheet \cite{viscoplasticPlates} and to combine the yield-stress intrusion model with an elastic fracture toughness at the tip.

The authors thank for the funding through the EarthFlow initiative at the University of Oslo. 


\bibliography{apssamp.bib}

\end{document}